\documentclass[aps,showpacs,preprintnumbers,amsmath,amssymb]{revtex4}
\usepackage{dcolumn}
\usepackage[dvips]{epsfig}
\usepackage{graphicx}
\usepackage{amssymb}
\usepackage{color}
\usepackage{enumerate}
\usepackage{subfigure}
\newcommand{\degree}{^\circ}

\begin{document}
\baselineskip=0.8 cm
%\preprint{example}
\title{Shadows of Bonnor black dihole by chaotic lensing }
%\altaffiliation{}
\author{Mingzhi Wang$^{1}$,  Songbai Chen$^{1,2,3}$\footnote{Corresponding author: csb3752@hunnu.edu.cn}, Jiliang
Jing$^{1,2,3}$ \footnote{jljing@hunnu.edu.cn}}
\affiliation{$ ^1$Institute of Physics and Department of Physics, Hunan
Normal University,  Changsha, Hunan 410081, People's Republic of
China \\ $ ^2$Key Laboratory of Low Dimensional Quantum Structures \\
and Quantum Control of Ministry of Education, Hunan Normal
University, Changsha, Hunan 410081, People's Republic of China\\
$ ^3$Synergetic Innovation Center for Quantum Effects and Applications,
Hunan Normal University, Changsha, Hunan 410081, People's Republic
of China}

\begin{abstract}
\baselineskip=0.6 cm
\begin{center}
{\bf Abstract}
\end{center}

 We have studied numerically the shadows of Bonnor black dihole through the technique of backward ray-tracing. The presence of magnetic dipole yields non-integrable photon motion, which affects sharply the shadow of the compact object. Our results show that there exists a critical value for the shadow. As the magnetic dipole parameter is less than the critical value, the shadow is a black disk, but as the magnetic dipole parameter is larger than the critical one, the shadow becomes a concave disk with eyebrows possessing a self-similar fractal structure. These behavior are very similar to those of the equal-mass and non-spinning Majumdar-Papapetrou binary black holes. However, we find that the two larger shadows and the smaller eyebrow-like shadows are joined together by the middle black zone for the Bonnor black dihole, which is different from that in the Majumdar-Papapetrou binary black holes spacetime where they are disconnected.
With the increase of magnetic dipole parameter, the middle black zone connecting the main shadows and the eyebrow-like shadows becomes narrow.
Our result show that the spacetime properties arising from the magnetic dipole yields the interesting patterns for the shadow casted by Bonnor black dihole.

\end{abstract}
\maketitle

\newpage
\section{Introduction}
A shadow is a two-dimensional dark region in the observer's sky corresponding to light rays that fall into an event horizon when propagated backwards in time. It is shown that the shape and size of the shadow carry the characteristic information of the geometry around the celestial body \cite{sha1,sha2,sha3}, which means that the shadow can be regarded as a useful tool to probe the nature of the celestial body and to check further various theories of gravity. The investigation \cite{sha2,sha3} indicate that the shadow is a perfect disk for a Schwarzschild black hole and it changes into an elongated silhouette for a rotating black hole due to its dragging effect. The cusp silhouette of shadow is found in the spacetime of a Kerr black hole with Proca hair \cite{fpos2} and of a Konoplya-Zhidenko rotating non-Kerr black hole \cite{sb10} as the black hole parameters lie in a certain range. Moreover, the shadow of a black hole with other characterizing  parameters have been studied recently \cite{sha4,sha5,sha6,sha7,sha9,sha10,sha11,sha12,sha13,sha14,sha14a,sha15,sha16,
sb1,sha17,sha19,shan1} (for details, see also a review \cite{shan1add}), which indicate that these  parameters bring the richer silhouettes for the shadows casted by black holes.

However, most of the above investigation have been focused only on the cases where the null geodesics are variable-separable and the corresponding dynamical systems are integrable. As the dynamical systems  are non-integrable, the motion of photons could be chaotic, which could lead to some novel features for the black hole shadow. Recently, it is shown that due to such chaotic lensing the multi-disconnect shadows with fractal structures emerge for a Kerr black hole with scalar hair \cite{sw,swo,astro,chaotic} or a binary black hole system \cite{binary,sha18}. The further analysis show that these novel patterns with fractal structures in shadows are determined by the non-planar bound orbits \cite{fpos2} and  the invariant phase space structures \cite{BI} for the photon motion in the black hole spacetimes. The similar analysis have also been done for the cases with ultra-compact object \cite{bstar1,bstar2}.

It is well known that there exist enormous magnetic
fields around large astrophysical black holes, especially
in the nucleus of galaxies \cite{Bm1,Bm2,Bm3,Bm4}. These strong magnetic fields could be
induced by currents in accretion disks near the supermassive galactic black holes. On the base of strong magnetic fields, there are substantial current theoretical models accounted for  black hole jets, which are one of the most spectacular astronomical events in the sky \cite{Blandford1,Blandford2,Punsly}. In general relativity, one of the most important solutions with magnetic fields is Ernst solution \cite{Ernst}, which describes the gravity of a black hole immersed in an external magnetic field. Interestingly, for an Ernst black hole,  the polar circumference for the event horizon increases with the magnetic field, while the equatorial circumference decreases. Bonnor's metric \cite{mmd1} is another important solution of the Einstein field equations in the vacuum, which describes  a static massive object with a dipole magnetic field in which two
static extremal magnetic black holes with charges of opposite signs
are situated symmetrically on the symmetry axis. For Bonnor black dihole spacetime, the area of the horizon is finite,  but the proper circumference of the horizon surface is zero. Especially, it is not a member of the
Weyl electromagnetic class and it can not reduce to Schwarzschild spacetime in the limit without magnetic dipole. The new properties of spacetime structure originating from magnetic dipole will lead to chaos in motion of particles \cite{mmd,mmd10,bbon1}. Since the shadow of black hole is determined by the propagation of light ray in the spacetime, it is expectable that the chaotic lensing caused by the new spacetime structure  will yields some new effects on the black hole shadow. Therefore, in this paper, we focus on studying the shadow of Bonnor black dihole  \cite{mmd1} and probe the effect of magnetic dipole parameter on the black hole shadow.

The paper is organized as follows. In Sec. II, we review briefly the metric of Bonnor black dihole and then analyze the propagation of light ray in this background. In Sec. III, we investigate the shadows casted by Bonnor black dihole. In Sec. IV, we discuss invariant phase space structures of photon motion and formation of the shadow casted by Bonnor black dihole. Finally, we present a summary.

\section{Spacetime of Bonnor black dihole and null geodesics}

Let us now to review briefly the spacetime of Bonnor black dihole. In 1960s, Bonnor obtained an exact solution \cite{mmd1} of Einstein-Maxwell equations which describes a static massive source carrying a magnetic dipole.
In the standard coordinates, the metric of this spacetime has a form \cite{mmd1}
\begin{eqnarray}
\label{xy}
ds^{2}= -\bigg(\frac{P}{Y}\bigg)^{2}dt^{2}+\frac{P^{2}Y^{2}}{Q^{3}Z}(dr^{2}+Zd\theta^{2})
+\frac{Y^{2}Z\sin^{2}\theta}{P^{2}}d\phi^{2},
\end{eqnarray}
where
\begin{equation}
P=r^{2}-2mr-b^{2}\cos^{2}\theta,\;\;Q=(r-m)^{2}-(m^{2}+b^{2})\cos^{2}\theta,
\;\;Y=r^{2}-b^{2}\cos^{2}\theta,\;\;Z=r^{2}-2mr-b^{2}.
\end{equation}
The corresponding vector potential $A_{\mu}$ is given by
\begin{eqnarray}
A_{\mu}= (0,0,0,\frac{2mbr\sin^{2}\theta}{P}),
\end{eqnarray}
where $\mu=0,1,2,3$ correspond to the element of $A_{\mu}$ associated with the coordinates $t, r, \theta, \phi$, respectively.
It is a static axially-symmetric solution characterized by two independent parameters $m$ and $b$, which are related to the total mass of Bonnor black dihole $M$ as $M=2m$ and to the magnetic
dipole moment $\mu$ as $\mu=2mb$. Obviously, this spacetime is asymptotically flat since as the polar coordinate $r$ approaches to infinity the metric tends to the Minkowski one. The event horizon of the spacetime (\ref{xy}) is the null hypersurface $f$ satisfied
\begin{eqnarray}
g^{\mu\nu}\frac{\partial f}{\partial x^{\mu}}\frac{\partial f}{\partial x^{\nu}}=0,
\end{eqnarray}
which yields
\begin{eqnarray}
r^{2}-2mr-b^{2}=0.
\end{eqnarray}
It is obvious that there exists only a horizon and the corresponding horizon radius is $r_h=m+\sqrt{m^2+b^2}$. The area of the horizon is $\mathcal{A}=16\pi m^2r^2_h/(m^2+b^2)$, but the proper circumference of the horizon surface is zero since $g_{\phi\phi}=0$ on the horizon. This implies that the $Z=0$ surface is not a regular horizon since there exists conical singularities at $r=r_h$. The singularity along the segment $r=r_h$ can be eliminated by selecting a proper period $\Delta\phi=2\pi[b^2/(m^2+b^2)]^2$, but such a choice yields a conical deficit running along the axes $\theta=0, \;\pi$, from the endpoints of the
dipole to infinity \cite{mmd101,mmd102}. The defects outside the dipole can be treated as open cosmic strings and then Bonnor black dihole is held apart by the cosmic strings that pull from its endpoints. Since the angular coordinate $\phi$ is periodic, an azimuthal curve $\gamma=\{t=Constant, r=Constant, \theta=Constant\}$ is a closed curve with invariant length $s^2_{\gamma}=g_{\phi\phi}(2\pi)^2$. And then the integral curve with $(t, r, \theta)$ fixed is closed timelike curve as $g_{\phi\phi}<0$. Thus, there exist closed timelike curves inside the horizon. However, the region outside the horizon is regular and there is no closed timelike curves. Moreover, the spacetime (\ref{xy}) possesses the complicated singular behaviour at $P=0$, $Q=0$ and
$Y=0$, but there is no singularity outside the horizon. As $b=0$, one can find that  it does not reduce to Schwarzschild spacetime, but to the Zipoy-Voorhees one with $\delta=2$ \cite{mmd12,mmd11}, which describes a monopole of mass $2m$
together with higher mass multipoles depended on the parameter $m$.
These special spacetime properties affect the propagation of photon and further changes  shadow of Bonnor black dihole(\ref{xy}).

The Hamiltonian of a photon motion along null geodesics in the spacetime (\ref{xy}) can be expressed as
\begin{equation}
\label{hami}
H(x,p)=\frac{1}{2}g^{\mu\nu}(x)p_{\mu}p_{\nu}=0.
\end{equation}
Since the metric functions in the spacetime (\ref{xy}) are independent of the coordinates $t$ and $\phi$, it is easy to obtain two conserved quantities $E$ and $L_{z}$ with the following forms
\begin{eqnarray}
\label{EL}
E=-p_{t}=-g_{00}\dot{t},\;\;\;\;\;\;\;\;\;\;\;\;\;\;
L_{z}=p_{\phi}=g_{33}\dot{\phi},
\end{eqnarray}
which correspond to the energy  and the $z$-component of the angular momentum
of photon moving in the background spacetime. With these two conserved quantities, we can obtain the equations of a photon motion along null geodesics
\begin{eqnarray}
\label{cdx}
\ddot{r}&=&-\frac{1}{2}\frac{\partial }{\partial r}\bigg[\ln\bigg(\frac{P^2Y^2}{Q^3Z}\bigg)\bigg]\dot{r}^{2}-\frac{\partial }{\partial \theta}\bigg[\ln\bigg(\frac{P^2Y^2}{Q^3Z}\bigg)\bigg]\dot{r}\dot{\theta}
+\frac{Z}{2}\frac{\partial }{\partial r}\bigg[\ln\bigg(\frac{P^2Y^2}{Q^3}\bigg)\bigg]\dot{\theta}^{2}\nonumber\\
&&-\frac{Q^3Z}{2}\bigg[\frac{E^2}{P^4}\frac{\partial}{\partial r}\ln\bigg(\frac{P^2}{Y^2}\bigg)-
\frac{L^2_z}{Y^4Z\sin\theta}\frac{\partial}{\partial r}\ln\bigg(\frac{Y^2Z\sin\theta}{P^2}\bigg)\bigg], \nonumber\\
\ddot{\theta}&=&\frac{1}{2Z}\frac{\partial }{\partial \theta}\bigg[\ln\bigg(\frac{P^2Y^2}{Q^3}\bigg)\bigg]\dot{r}^{2}-\frac{\partial }{\partial r}\bigg[\ln\bigg(\frac{P^2Y^2}{Q^3Z}\bigg)\bigg]\dot{r}\dot{\theta}
+\frac{1}{2}\frac{\partial }{\partial \theta}\bigg[\ln\bigg(\frac{P^2Y^2}{Q^3}\bigg)\bigg]\dot{\theta}^{2}\nonumber\\
&&-\frac{Q^3}{2}\bigg[\frac{E^2}{P^4}\frac{\partial}{\partial \theta}\ln\bigg(\frac{P^2}{Y^2}\bigg)-
\frac{L^2_z}{Y^4Z\sin\theta}\frac{\partial}{\partial \theta}\ln\bigg(\frac{Y^2Z\sin\theta}{P^2}\bigg)\bigg],
\end{eqnarray}
with the constraint condition
\begin{eqnarray}
\label{lglr}
H=\frac{1}{2}\bigg(\frac{Q^{3}Z}{P^{2}Y^{2}}p_{r}^{2}+\frac{Q^{3}}{P^{2}Y^{2}}p_{\theta}^{2}
+V\bigg)=0,
\end{eqnarray}
where  $p_{r}$ and $p_{\theta}$ are the components of momentum of the photon $p_{r}=g_{11}\dot{r}$ and $p_{\theta}=g_{22}\dot{\theta}$. $V$ is the effective potential with a form
\begin{eqnarray}
\label{vv}
V=-(\frac{Y}{P})^{2}E^{2}+\frac{P^{2}}{Y^{2}Z\sin^{2}\theta}L_{z}^{2}.
\end{eqnarray}
Obviously, in the case with magnetic dipole (i.e.,$b\neq0$), we find that the equations of motion (\ref{cdx}) and (\ref{lglr}) can not be variable-separable and the corresponding dynamical system is non-integrable because it admits
only two integrals of motion $E$ and $L_z$. This implies that the motion of the photon could be chaotic in the spacetime (\ref{xy}), which will bring some new features for the shadow casted by Bonnor black dihole.

\section{Shadow casted by Bonnor black dihole}

In this section, we will study the shadow casted by Bonnor black dihole with
the method called ``backward ray-tracing" \cite{sw,swo,astro,chaotic} in which the light rays are assumed to evolve from the observer backward in time. In this method, we must solve numerically the null geodesic equations (\ref{EL}) and (\ref{cdx}) for each pixel in the final image with the corresponding initial condition. The image of shadow in observer's sky is composed of the pixels corresponding to the light rays falling down into the horizon of black hole.

Since the spacetime of Bonnor black dihole (\ref{xy}) is asymptotic flat, we can define the same observer's sky at spatial infinite as in the usual static cases. The observer basis $\{e_{\hat{t}},e_{\hat{r}},e_{\hat{\theta}},e_{\hat{\phi}}\}$ can be expanded in the coordinate basis $\{ \partial_t,\partial_r,\partial_{ \theta},\partial_{\phi} \}$  as a form \cite{sw,swo,astro,chaotic}
\begin{eqnarray}
\label{zbbh}
e_{\hat{\mu}}=e^{\nu}_{\hat{\mu}} \partial_{\nu},
\end{eqnarray}
where $e^{\nu}_{\hat{\mu}}$ satisfies $g_{\mu\nu}e^{\mu}_{\hat{\alpha}}e^{\nu}_{\hat{\beta}}
=\eta_{\hat{\alpha}\hat{\beta}}$, and $\eta_{\hat{\alpha}\hat{\beta}}$ is the usual Minkowski metric. For a static spacetime, it is convenient to choice a decomposition
\begin{eqnarray}
\label{zbbh1}
e^{\nu}_{\hat{\mu}}=\left(\begin{array}{cccc}
\zeta&0&0&0\\
0&A^r&0&0\\
0&0&A^{\theta}&0\\
0&0&0&A^{\phi}
\end{array}\right),
\end{eqnarray}
where $\zeta$,  $A^r$, $A^{\theta}$, and $A^{\phi}$ are real coefficients.
From the Minkowski normalization, one can find that the observer basis obey
\begin{eqnarray}
e_{\hat{\mu}}e^{\hat{\nu}}=\delta_{\hat{\mu}}^{\hat{\nu}}.
\end{eqnarray}
Therefore, we have
\begin{eqnarray}
\label{xs}
\zeta=\frac{1}{\sqrt{-g_{00}}},\;\;\;\;\;\;\;\;\;\;\;\;\;\; A^r=\frac{1}{\sqrt{g_{11}}},\;\;\;\;\;\;\;\;\;\;\;\;\;\;\;\;
A^{\theta}=\frac{1}{\sqrt{g_{22}}},\;\;\;\;\;\;\;\;\;\;\;\;\;\;\;
A^{\phi}=\frac{1}{\sqrt{g_{33}}},
\end{eqnarray}
and then the locally measured four-momentum $p^{\hat{\mu}}$ of a photon can be obtained by the projection of its four-momentum $p^{\mu}$  onto $e_{\hat{\mu}}$,
\begin{eqnarray}
\label{dl}
p^{\hat{t}}=-p_{\hat{t}}=-e^{\nu}_{\hat{t}} p_{\nu},\;\;\;\;\;\;\;\;\;
\;\;\;\;\;\;\;\;\;\;\;p^{\hat{i}}=p_{\hat{i}}=e^{\nu}_{\hat{i}} p_{\nu},
\end{eqnarray}
In the spacetime of Bonnor black dihole (\ref{xy}),  the locally measured four-momentum $p^{\hat{\mu}}$ can be further written as
\begin{eqnarray}
\label{smjt}
p^{\hat{t}}&=&\frac{1}{\sqrt{-g_{00}}}E,\;\;\;\;\;\;\;\;\;\;\;\;\;\;\;\;\;\;\;\;
p^{\hat{r}}=\frac{1}{\sqrt{g_{11}}}p_{r} ,\nonumber\\
p^{\hat{\theta}}&=&\frac{1}{\sqrt{g_{22}}}p_{\theta},
\;\;\;\;\;\;\;\;\;\;\;\;\;\;\;\;\;\;\;\;\;\;
p^{\hat{\phi}}=\frac{1}{\sqrt{g_{33}}}L_z.
\end{eqnarray}
After some similar operations in Refs.\cite{sw,swo,astro,chaotic}, we can obtain the position of photon's image in observer's sky \cite{sb10}
\begin{eqnarray}
\label{xd1}
x&=&-r_{obs}\frac{p^{\hat{\phi}}}{p^{\hat{r}}}=-r_{obs}\frac{L_{z}}{\sqrt{g_{11} g_{33}}\dot{r}}, \nonumber\\
y&=&r_{obs}\frac{p^{\hat{\theta}}}{p^{\hat{r}}}=
r_{obs}\frac{\sqrt{g_{22}}\dot{\theta}}{\sqrt{g_{11}}\dot{r}}.
\end{eqnarray}
\begin{figure}[ht]
\center{\includegraphics[width=6cm ]{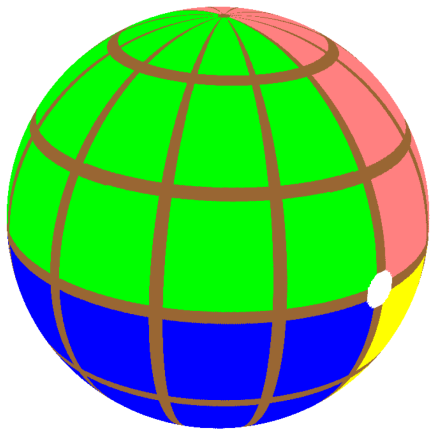}
\caption{The sphere light source marked by four different color quadrants and the brown grids of longitude and latitude. The white reference spot lies at intersection of the four colored quadrants.}
\label{gy}}
\end{figure}
\begin{figure}
\center{\includegraphics[width=14cm ]{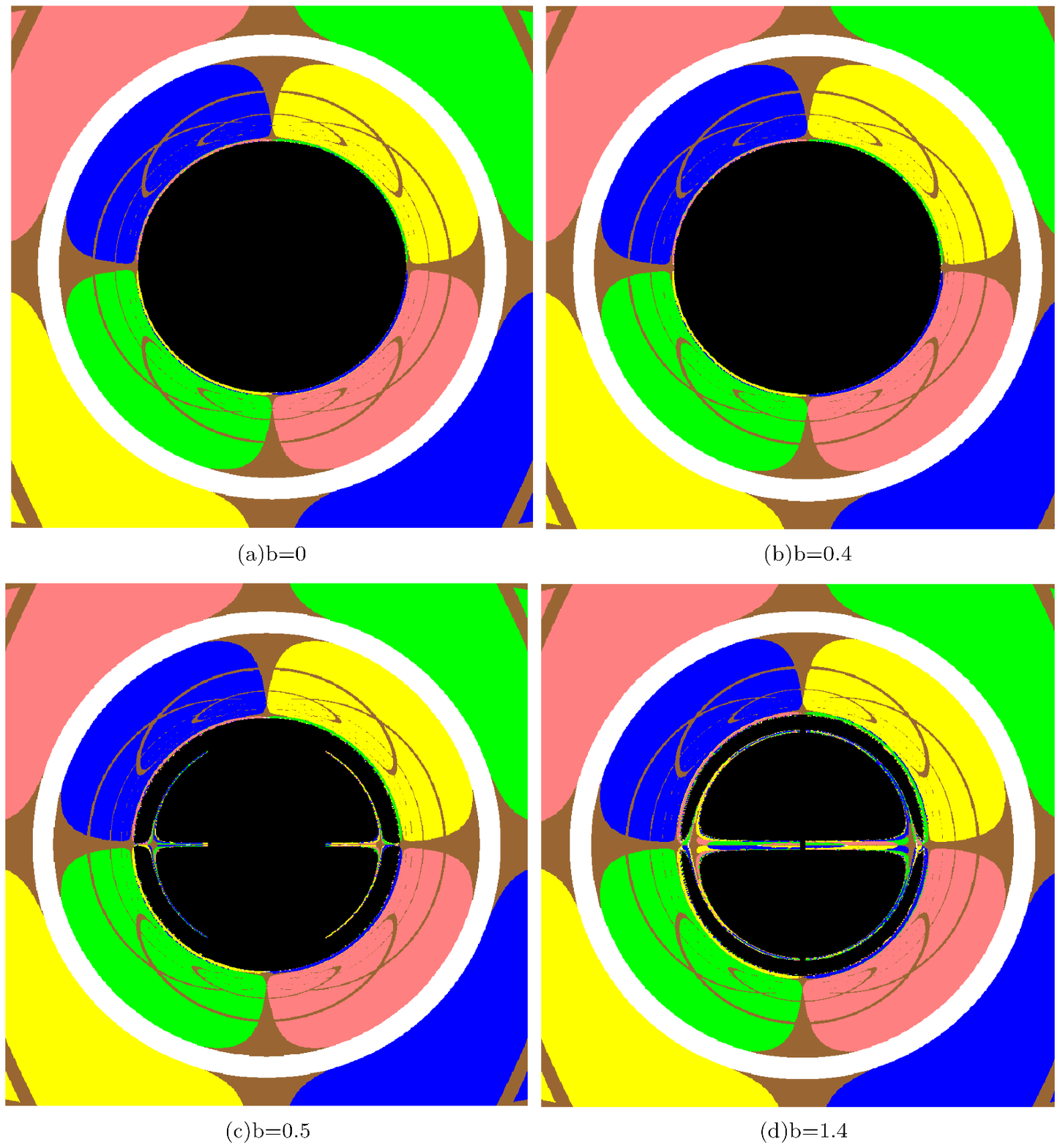}
\caption{The shadow casted by Bonnor black dihole (\ref{xy}) with different $b$. Here we set $m=1$ and the observer is set at $r_{obs}=30m$ with the inclination angle $\theta_{0}=90\degree$.}
\label{shb}}
\end{figure}
Following the way done in \cite{sw,swo,astro,chaotic,binary,sha18}, one can divide celestial sphere into four quadrants marked with a different color (green, blue, red and yellow as shown in FIG.\ref{gy}). The grid of longitude and latitude lines are
marked with adjacent brown lines separated by $10^\circ$. The observer is placed
off-centre within the celestial sphere at some a real radius
$r_{obs}$. For the sake of simplify, it is placed at the intersection of the four colored quadrants on the celestial sphere, i.e., $r_{obs}=r_{sphere}$, which is not shown in Fig.\ref{gy}.
The white reference spot in Fig.\ref{gy} lies at the other intersection of the four colored quadrants, which could provide a direct demonstration of Einstein ring \cite{sw,swo,astro,chaotic,binary,sha18}. We can integrate these null
geodesics with different initial conditions until they either reach a point on the celestial
sphere or they fall into the horizon of the compact object and the latter defines the shadow.
\begin{figure}
\begin{center}
\includegraphics[width=14cm ]{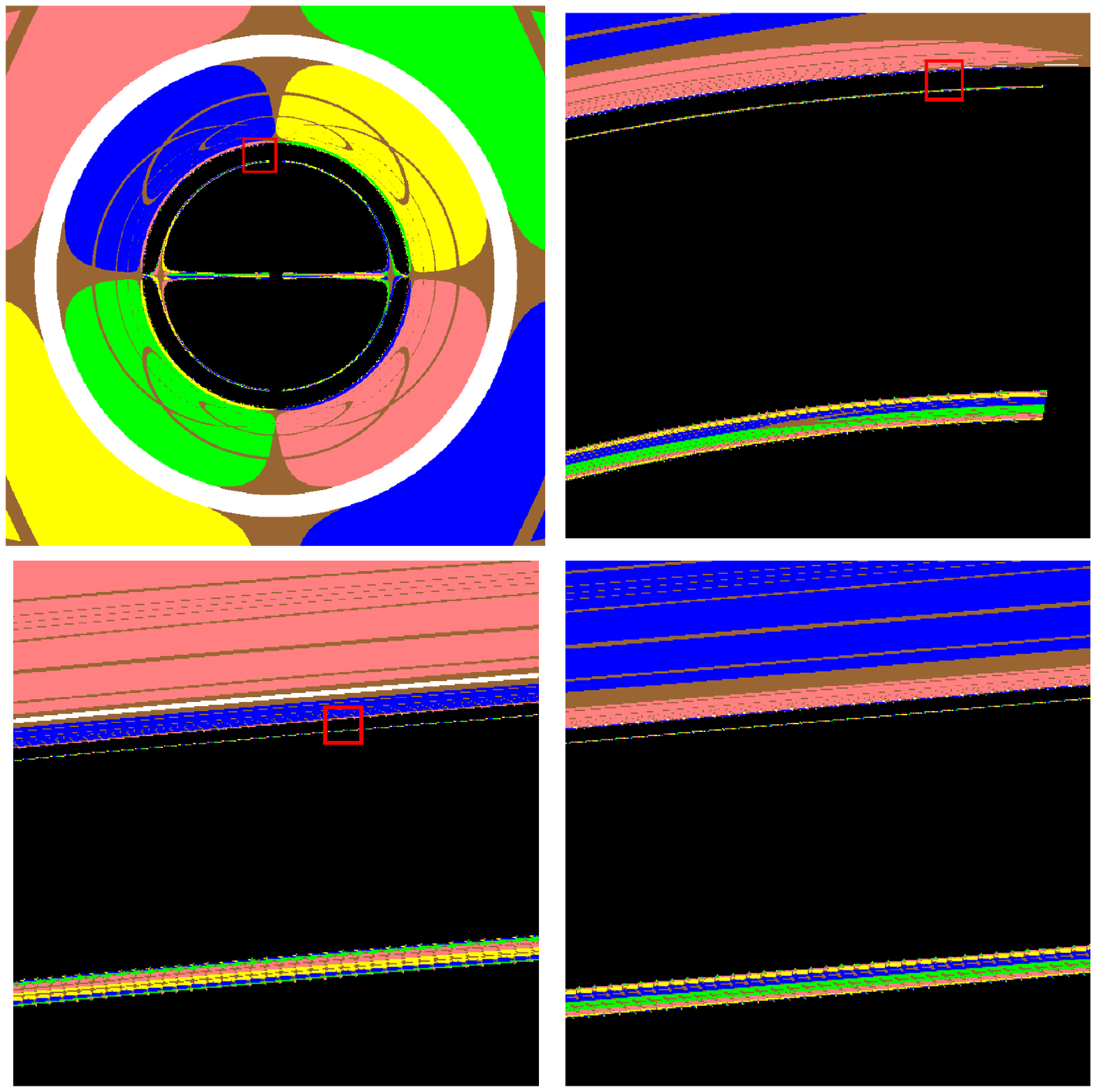}
\end{center}
\caption{The fractal structure in the shadow of Bonnor black dihole (\ref{xy}) for fixed $b=1.0$. Here we set $m=1$ and the observer is set at $r_{obs}=30m$ with the inclination angle $\theta_{0}=90\degree$. }
\label{fx}
\end{figure}
In Fig. \ref{shb}, we present the shadow casted by Bonnor black dihole (\ref{xy}) with different $b$. Here we set $m=1$ and the observer is set at $r_{obs}=30m$ with the inclination angle $\theta_{0}=90\degree$. Our numerical results show that there exists a critical value $b_c\sim 0.404$ for the shadow. As $b<b_c$, we find that the shadow is a black disk, which is similar to those in the usual static compact object spacetimes with horizon. Moreover, we find that the size of the shadow decreases with the parameter $b$  in this case. However, for the case $b>b_c$,  there exist two anchor-like bright zones imbedded symmetrically in the black disk shadow so that the shadow looks like a concave disk with four larger eyebrows, which are shown in Fig. \ref{shb} (c)-(d). The eyebrow-like
features of shadow are also found in Refs.\cite{sw,swo,astro,chaotic,binary,sha18}.  Actually, many other smaller eyebrow-like shadows can be detected in two anchor-like bright zones as shown in Fig.\ref{fx}. This hints that the shadow possess  a self-similar fractal structure, which is caused by chaotic lensing. It is an interesting property of shadows, which is qualitatively  different from those in the spacetimes where the equations of motion are variable-separable and the corresponding dynamical system is integrable. With the increase of magnetic dipole parameter $b$,  the eyebrows becomes long and the fractal structure become more rich. Moreover, we find that the two anchor-like bright zones increase with the parameter $b$, but for arbitrary $b$, two anchor-like bright zones are disconnected since they are always separated by a black region. In other words, for Bonnor black dihole, the two larger shadows and the smaller eyebrow-like shadows are joined together by the middle black zone.
Moreover, the white circle in Figs. \ref{shb} and \ref{fx} denote Einstein ring, which are consistent with the prediction of multiple images of a source due to gravitational lensing.

\section{Invariant phase space structures and formation of shadow casted by Bonnor black dihole}

In this section, we will discuss the formation of the shadow casted by Bonnor black dihole through analysing the invariant phase space structures as in Ref. \cite{BI}.
The invariant phase space structures including fixed points, periodic orbits and invariant manifolds, are one of important features for dynamical systems, which are applied extensively in the design of space trajectory for various of spacecrafts, such as, a low energy transfer from the Earth to the Moon and a ``Petit Grand Tour" of the moons of Jupiter  \cite{BI17,BI18,BI19,BI20,BI22}.
Recent investigations \cite{BI} show that these invariant structures play an important role in the emergence of black hole shadows.

For the spacetime of Bonnor black dihole (\ref{xy}), the fixed point $x_0=(r_{0},\theta_{0},0,0)$ in phase space $(r,\theta,p_r,p_{\theta})$ satisfies the condition
\begin{eqnarray}
\label{bdd}
\dot{x}^{\mu}=\frac{\partial H}{\partial p_{\mu}}=0,\;\;\;\;\;\;\;\;\;\;\;\;\;\;
\dot{p}_{\mu}=-\frac{\partial H}{\partial x^{\mu}}=0,
\end{eqnarray}
which means
\begin{eqnarray}
\label{bdd1}
 V\bigg|_{r_{0},\theta_{0}}=0,\;\;\;\;\;\;\;\;\;\;\;\;\;\;\frac{\partial V}{\partial r}\bigg|_{r_{0},\theta_{0}}=0,\;\;\;\;\;\;\;\;\;\;\;\;\;\;
\frac{\partial V}{\partial \theta}\bigg|_{r_{0},\theta_{0}}=0.
\end{eqnarray}
The local stability of the fixed point $x_0=$($r_{0},\theta_{0},0,0$) can be obtained by
linearizing the equations (\ref{bdd})
\begin{eqnarray}
\label{xxh}
\mathbf{\dot{X}}=J\mathbf{X},
\end{eqnarray}
where $\mathbf{X}=(\tilde{x}^{\mu},\tilde{p}_{\mu})$ and $J$ is the Jacobian.
The circular photon orbits in the  equatorial plane named light rings are fixed points of the dynamics for the  photon motion \cite{BI,fpos2}.
After linearizing the equations (\ref{bdd}) near the fixed point $(r_{0},\pi/2,0,0)$ and setting $m=1$, we obtain the Jacobian
\begin{equation}
\label{jjj}
J=\left[ \begin{array}{cccc}
0 & 0 & 2A & 0 \\
0 & 0 & 0 & 2B \\
-2C & 0 & 0 & 0 \\
0 & -2D & 0 & 0
\end{array}
\right],
\end{equation}
with
\begin{eqnarray}
\label{jjt}
A&=&\frac{(r_{0}-1)^{6}(r_{0}^{2}-2r_{0}-b^{2})}{r_{0}^{6}(r_{0}-2)^{2}},\\ \nonumber
B&=&\frac{(r_{0}-1)^{6}}{r_{0}^{6}(r_{0}-2)^{2}},\\ \nonumber
C&=&\frac{\eta^{2}[3r_{0}^{2}(r_{0}-4)(r_{0}-2)^{3}+b^{2}r_{0}
(r_{0}-2)^{2}(16+r_{0})-4b^{4}(r_{0}-3)]}{r_{0}^{4}
(r_{0}^{2}-2r_{0}-b^{2})^3}-4\frac{r_{0}+1}{(r_{0}-2)^{4}},\\ \nonumber
D&=&\frac{\eta^{2}(r_{0}-2)(r_{0}^{3}-2r_{0}^{2}-4b^{2})}{r_{0}^{4}
(r_{0}^{2}-2r_{0}-b^{2})}-\frac{4b^{2}}{(r_{0}-2)^{3}},\\ \nonumber
r_{0}&=&\frac{1}{3}\bigg[(3\sqrt{3}\sqrt{108b^4-112b^2-225}-54b^2+28)^{1/3}+7+
\frac{19}{(3\sqrt{3}\sqrt{108b^4-112b^2-225}-54b^2+28)^{1/3}}\bigg],
\end{eqnarray}
where $\eta\equiv L_z/E$.
\begin{figure}[ht]
\center{\includegraphics[width=9cm ]{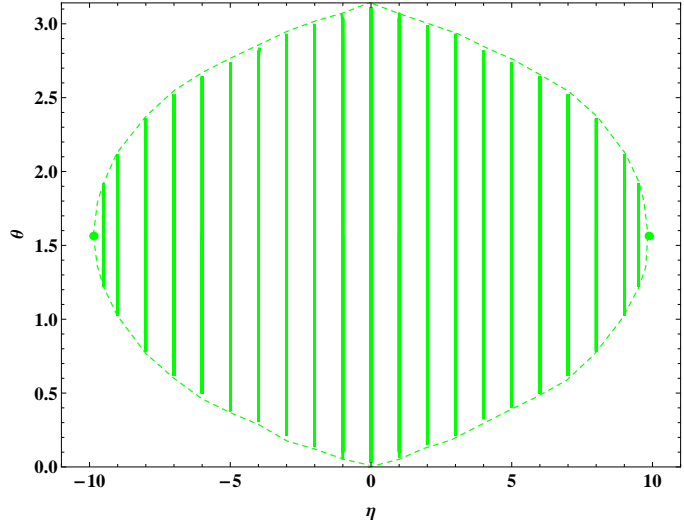}
\caption{Light rings (dots) and the corresponding family of periodic Lyapunov orbits (solid line) in the spacetime of Bonnor black dihole with $b=1.4$. Here we set $m=1$.}
\label{zqt}}
\end{figure}

Let us now adopt the case $m=1$ and $b=1.4$ as an
example to analyse the formation of the shadow of Bonnor black dihole (\ref{xy}) which is shown in Fig.\ref{shb} (d). In this special case, we find that there exist two fixed points. Their positions in phase space are overlapped  at ($4.07,\pi/2,0,0$), but their impact parameters are  $\eta_1=-9.83$ and $\eta_2=9.83$, respectively. The special distribution of two fixed points is attributed to that the considered magnetic dipole spacetime (\ref{xy}) is a non-rotating spacetime. The eigenvalues of the Jacobian (\ref{jjj}) are  $\pm \lambda$, $\pm \nu i$, where $\lambda=0.46$ and $\nu=0.60$. According to Lyapunov central theorem, we know  that each purely imaginary eigenvalue gives rise to a one parameter family $\gamma_{\epsilon}$ of periodic orbits, which is the so-called Lyapunov family \cite{BI} and the orbit $\gamma_{\epsilon}$ collapses into the fixed point as $\epsilon\rightarrow0$. We show Lyapunov family for the above fixed points (light rings) in Fig. \ref{zqt}.  The two thick dots represent the two light rings, and the solid lines denote a family of periodic Lyapunov orbits arising from these two light rings.
These periodic orbits can be parameterized by impact parameter $\eta$ in an interval $[-9.83,9.83]$. All of these periodic Lyapunov orbits are nearly spherical orbits with radius $r=4.07$, which are responsible for determining the boundary of shadow of Bonnor black dihole as in Refs. \cite{fpos2,BI}.
\begin{figure}[ht]
\center{\includegraphics[width=8cm ]{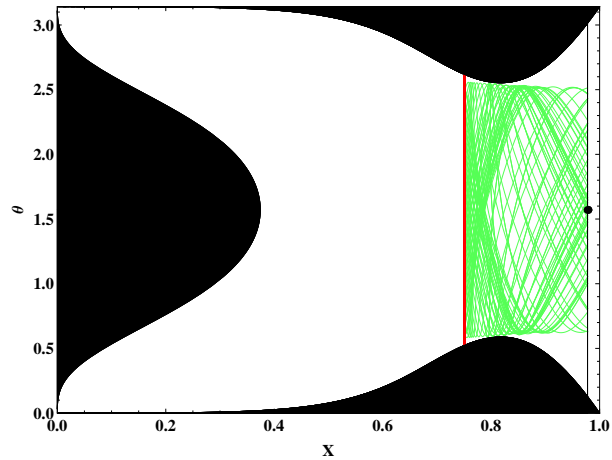}
\caption{Projection of the unstable invariant manifolds (green lines) associated with the periodic orbit for $\eta$ =-6 (red line). The dark region are the forbid region of photon and the black dot represents the position of observer.}
\label{sn}}
\end{figure}
The positive ( negative ) real eigenvalue $\pm \lambda$ suggests that there is a unstable ( stable ) invariant manifold, in which points exponentially approach the fixed point in backward ( forward ) time. For each Lyapunov orbit, its corresponding  invariant manifolds are two dimensional surfaces forming tubes in the three dimensional reduced phase space $(r; \theta; p_{\theta})$. In Fig. \ref{sn}, we show a projection of the unstable invariant manifolds associated with the periodic orbits for $\eta=-6$ in the plane ($X,\theta$), where $X$ is a compactified radial coordinate defined as $X=\sqrt{r^{2}-r_{h}^{2}}/(1+\sqrt{r^{2}-r_{h}^{2}})$ \cite{BI}. The orbits inside the unstable invariant manifold tube can reach the horizon of Bonnor black dihole.
Moreover, we note the periodic orbit touched the boundary of the black region approaches
perpendicularly to the boundary $V(r,\theta)= 0$ as in Ref.\cite{binary}.

\begin{figure}
\includegraphics[width=13cm ]{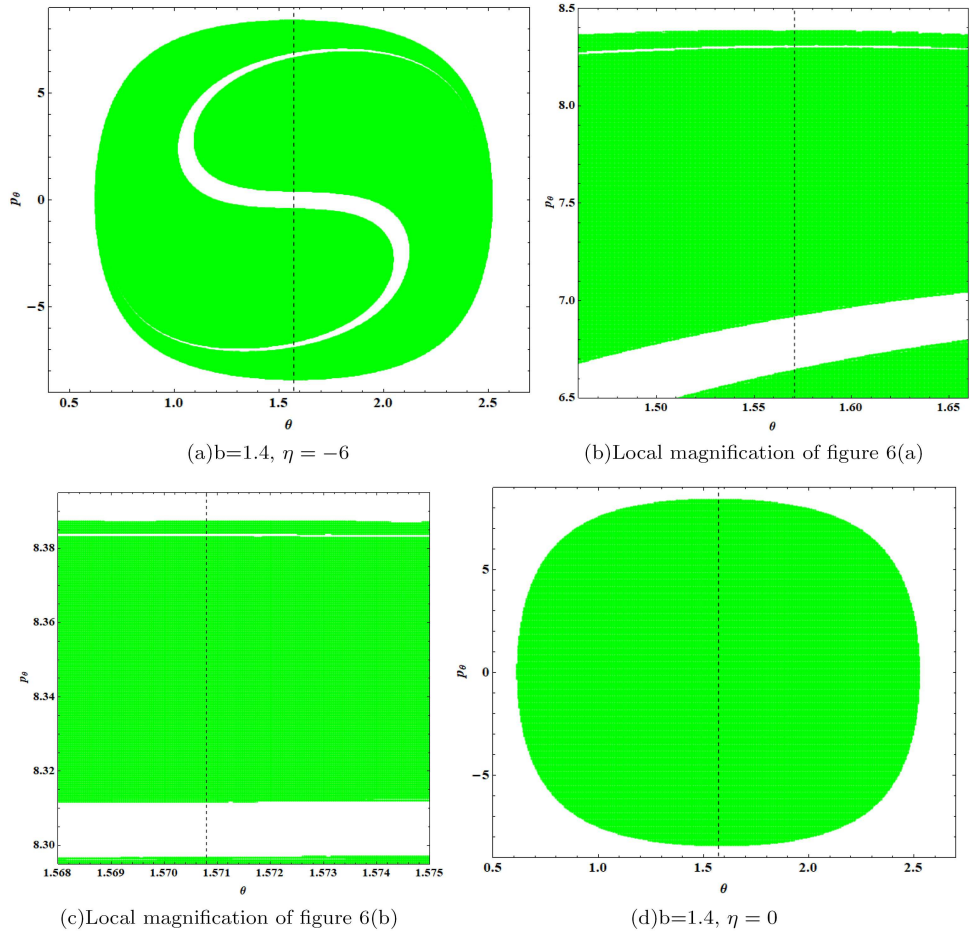}
\caption{The Poincar\'{e} section at $r=r_{obs}$ for the unstable manifolds (green) of Lyapunov orbits in the spacetime of Bonnor black dihole (\ref{xy}) with $b=1.4$. The figures (a)-(c) show the fractal-like structure for $\eta=-6$ and the figures (d) is $\eta=0$. Here we set $m=1$. }
\label{pjl}
\end{figure}

In order to probe the shape of the invariant manifolds  as in Ref.\cite{BI}, we present in Fig.\ref{pjl} the Poincar\'{e} section in the plane ($\theta, p_{\theta}$) for the unstable manifolds of Lyapunov orbits at the observers radial position with $\eta=-6$ and $\eta=0$.
All photons starting within the green regions always move only in the unstable manifold tube. Moreover, we also note that there exist some white regions which corresponds to that photons move outside the unstable manifolds. In Fig.\ref{pjl},  the intersection of the dashed line $\theta=\frac{\pi}{2}$ with these manifolds denotes the trajectories which can be detected by the observer on the equatorial plane. This can be generalized to the cases with other values of $\theta$.
\begin{figure}
\center{\includegraphics[width=8cm ]{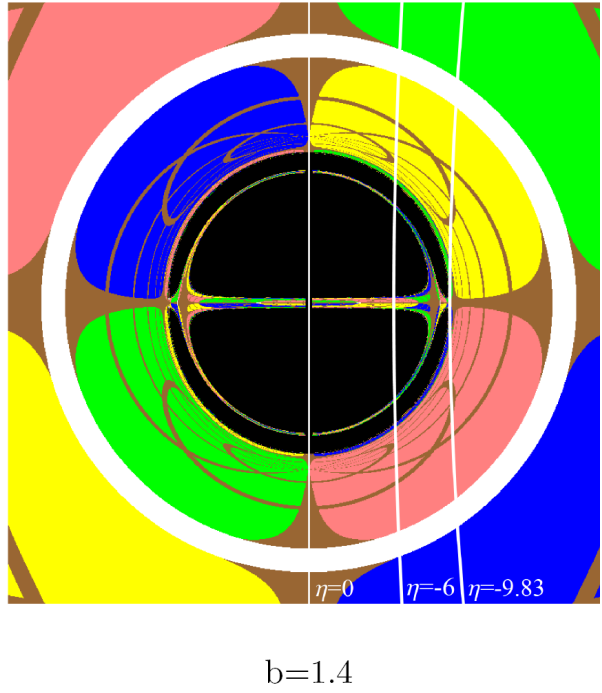}}
\caption{Intersections of the unstable manifolds  with the
image plane for the lines with the constant $\eta=-9.83$, $\eta=-6$, and $\eta=0$ in the spacetime of Bonnor black dihole (\ref{xy}) with $b=1.4$.}
\label{jx}
\end{figure}
Actually, these intersection points also determine the positions of the photons with a certain  angular momentum on the image plane. In Fig. \ref{jx}, we present the lensing image marking the intersection points for fixed $\eta=-9.83$, $\eta=-6$, and $\eta=0$. The boundary of the shadow of Bonnor black dihole are determined entirely by the intersection points deriving from these fixed points. The anchor-like bright zones in Fig. \ref{shb} (d) are originated from the top, middle and bottom parts of the $S-$shape white region in the Poincar\'{e} section ( see in Fig.\ref{pjl} (a)) and the fractal-like structure shown in Fig.\ref{pjl} (a)-(c) is responsible for the fractal shadow structure in Fig.3. For the case $\eta=0$, there is no white region in the  Poincar\'{e} section ( see in Fig.\ref{pjl} (d)), which is responsible for that two anchor-like bright zones are  separated by the black shadow in the middle regions in Fig. \ref{shb} (d).
\begin{figure}
\center{\includegraphics[width=9cm ]{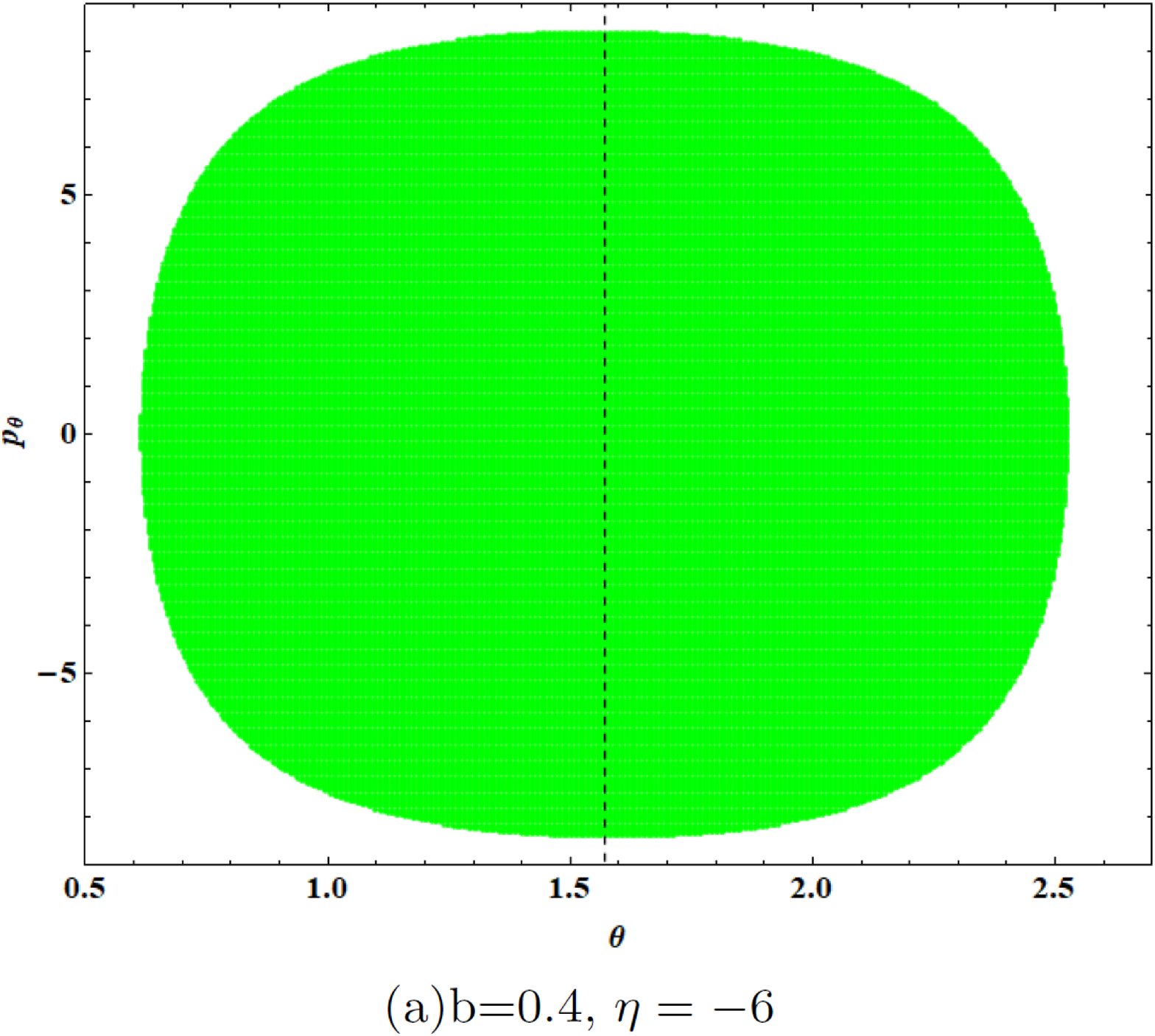} \includegraphics[width=7cm] {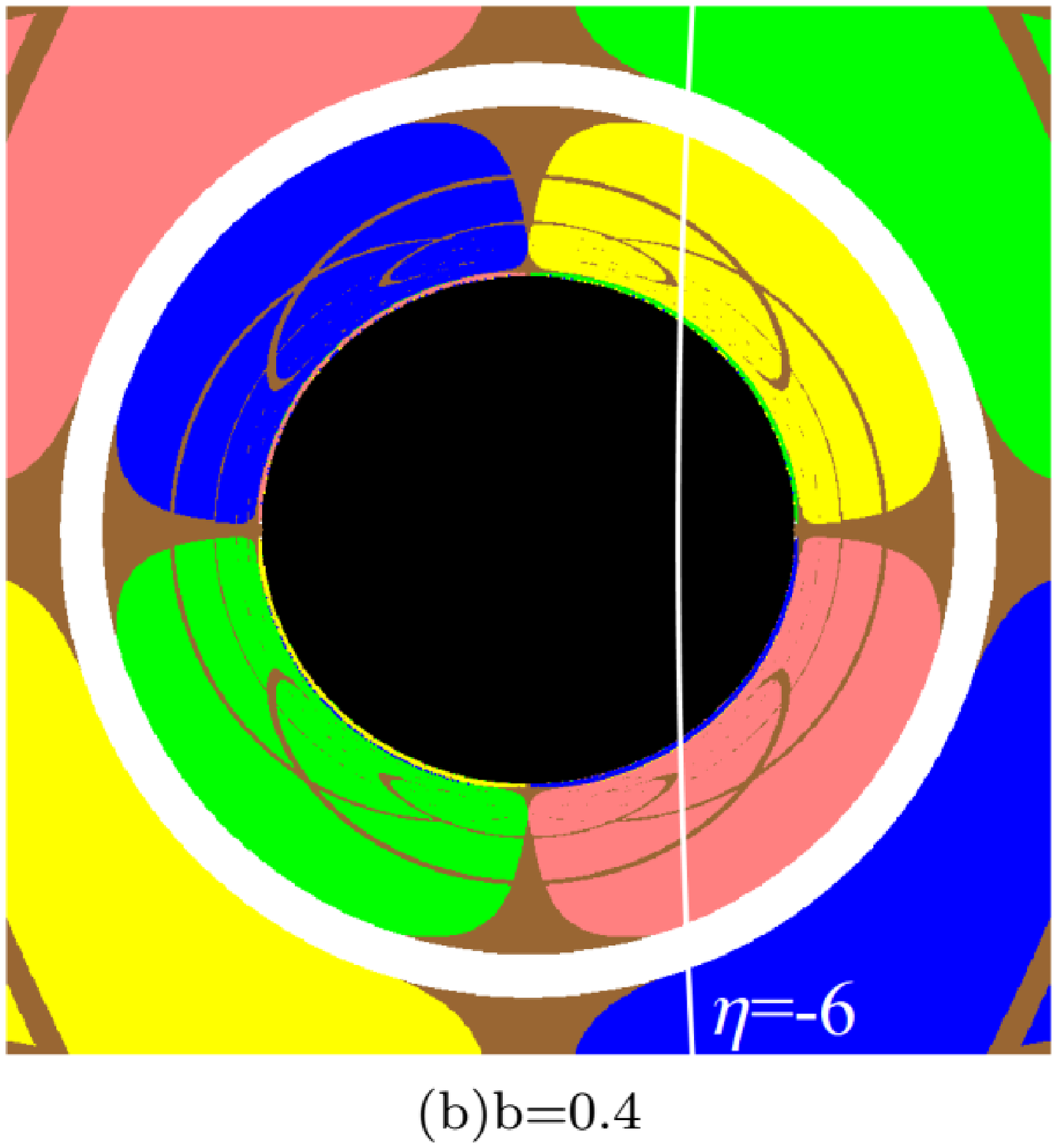}}
\caption{The Poincar\'{e} section (the left) and the intersections of the unstable manifolds  with the image plane (the right) for $\eta=-6$  in the spacetime of Bonnor black dihole (\ref{xy}) with $b=0.4$. }
\label{jx04}
\end{figure}
In order to make a comparison, in Fig. \ref{jx04}, we also plot the Poincar\'{e} section and the intersections of the unstable manifolds  with the image plane for $\eta=-6$ in the spacetime of Bonnor black dihole (\ref{xy}) with $b=0.4$. Obviously, there is no white region in the Poincar\'{e} section, which is consistent with that the shadow of Bonnor black dihole is a black disk and there exist no bright zones in the shadow in this case.

Finally, we make a comparison between the shadows casted by the equal-mass and non-spinning Majumdar-Papapetrou binary black holes \cite{binary,sha18} and by Bonnor black dihole (\ref{xy}). In Fig.9, we present the shadow for the Majumdar-Papapetrou binary black holes \cite{binary,sha18} with two equal-mass black
holes separated by the parameter $a=0.5$, $a=1$ and $a=2$ ( see figures (a)-(c)) and for
the cases of Bonnor black dihole (\ref{xy}) separated by the parameter $b=0.5$, $b=1$ and $b=2$ ( see figures (d)-(e)). From Fig.9, one can find that the shadows of Bonnor black dihole possess some properties closely resembling those of Majumdar-Papapetrou binary black holes, which is understandable since there exists the similar black hole configurations in both cases. However, there exists the essential difference in the shadows for the chosen parameter in these two cases. From Fig.9, we find that the two larger shadows and the smaller eyebrow-like shadows are joined together by the middle black zone for Bonnor black dihole, but they are disconnected in the case of the equal-mass and non-spinning Majumdar-Papapetrou binary black holes \cite{binary,sha18}.
\begin{figure}
\includegraphics[width=10cm ]{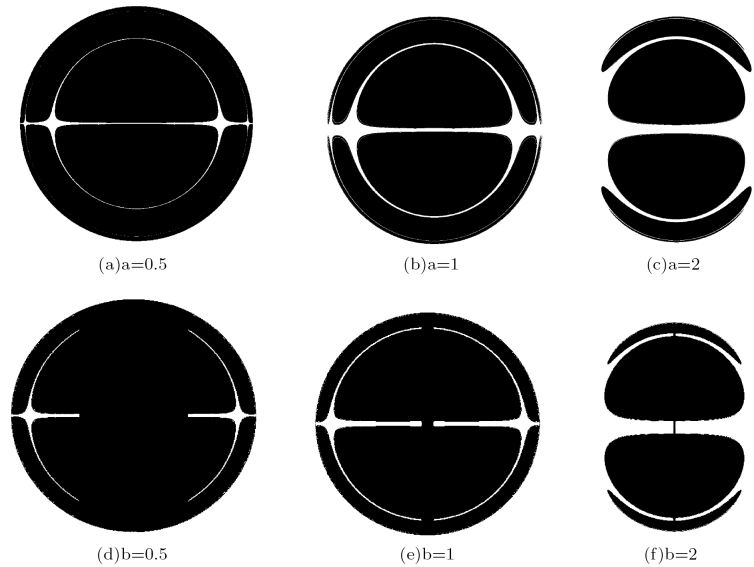}
\caption{The comparison between the shadows of Majumdar-Papapetrou binary black holes  and of   Bonnor black dihole (\ref{xy}). Figures (a), (b) and (c) correspond to the Majumdar-Papapetrou binary case \cite{binary,sha18} with two equal-mass black holes separated by the parameter $a=0.5$, $a=1$ and $a=2$, respectively. Figures (d),(e) and (f)
denote the shadow for the cases of Bonnor black dihole (\ref{xy}) separated by the parameter $b=0.5$, $b=1$ and $b=2$, respectively. Here we set the inclination angle of observer $\theta_{0}=90\degree$ and $m=1$. }
\end{figure}
Moreover, with
the increase of magnetic dipole parameter, we find that the middle black zone connecting the main shadows and the eyebrow-like shadows becomes narrow for Bonnor black dihole. From the previous discussion, we know that due to the existence of singularity on the symmetric axis, Bonnor black dihole is held apart by the cosmic string with tension
$\mu=\frac{1}{4}[1-b^4/(m^2+b^2)^2]$ \cite{mmd101,mmd102}, which decreases with the parameter $b$. Therefore, we can obtain that the middle black zone increases with the tension of the cosmic string. This behavior is consistent with that in the case of
a Kerr black hole pierced by a cosmic string in which the size of black hole shadow increases with the string tension \cite{sha14a}. Therefore, the appearance of the middle black zone in the shadow of Bonnor black dihole can be attributed to
the existence of the conical singularity on the symmetric axis in the background spacetime.
In the case of Majumdar-Papapetrou binary black holes \cite{binary,sha18}, there is no such conical singularity since the configuration is supported by the balance between the gravitational force and the Coulomb force. Thus, the difference of
the shadow shape in these two spacetimes is caused by the existence
of singularity on the symmetric axis in Bonnor's spacetime.

\section{Summary}

In this paper we have studied the shadows of black dihole described by Bonnor's exact solution of Einstein-Maxwell equations. The presence of magnetic dipole yields that the equation of photon motion can not be variable-separable and the corresponding dynamical system is non-integrable. With the technique of backward ray-tracing,
we present numerically the shadow of Bonnor black dihole.
For the smaller magnetic dipole parameter $b$, the shadow is a black disk as in the usual static compact object spacetimes with horizon. The size of shadow decreases with the parameter $b$. For the larger magnetic dipole parameter $b$, we find that there exist two anchor-like bright zones imbedded symmetrically in the black disk shadow so that the shadow looks like a concave disk with four large eyebrows. The anchor-like bright zones increase and the eyebrows become long with the increase of $b$. Moreover, many other smaller eyebrow-like shadows can be detected in two anchor-like bright zones and the shadow possess a self-similar fractal structure, which is caused by chaotic lensing. This interesting property of shadows is qualitatively  different from those in the spacetimes in which the equations of motion are variable-separable and the corresponding dynamical system is integrable.
Finally, we analyse the invariant manifolds of certain Lyapunov orbits near the fixed point and discuss further the formation of the shadow of Bonnor black dihole, which indicates that all of the structures in the shadow originate naturally from the dynamics near fixed points. Our result show that the spacetime properties arising from the magnetic dipole yields some interesting patterns for the shadow casted by Bonnor black dihole.

Comparing with that in the case of Majumdar-Papapetrou binary black holes, we find that the two larger shadows and the smaller eyebrow-like shadows are joined together by the middle black zone for Bonnor black dihole, but they are disconnected in the Majumdar-Papapetrou one.
The appearance of the middle black zone in the shadow of Bonnor black dihole can be attributed to the existence of the conical singularity on the symmetric axis in the background spacetime.
It is of interest to study the effects of such conical singularity on the Lyapunov orbits and the shadow edge etc. Work in this direction will be reported in the
future \cite{wchen}.

\section{\bf Acknowledgments}

We wish to thank anonymous referees very much for their useful comments and suggestions, which improve our paper considerably. This work was partially supported by the Scientific Research
Fund of Hunan Provincial Education Department Grant
No. 17A124. J. Jing's work was partially supported by
the National Natural Science Foundation of China under
Grant No. 11475061.

\end{document}